\documentclass[prl,superscriptaddress,preprint]{revtex4}
\usepackage{dcolumn}
\usepackage{bm}
\usepackage{amssymb}
\usepackage{amsmath}
\usepackage{amsfonts}
\usepackage[english]{babel}
\usepackage{amsmath}
\usepackage{graphicx}
\usepackage{dcolumn}
\usepackage{bm}
\usepackage{hyperref}
\usepackage{bigints}

\RequirePackage[cdot,textstyle,amssymb]{SIunits}%
\RequirePackage{units}
\usepackage[usenames, dvipsnames]{color}

\usepackage[]{graphicx}

\usepackage{epstopdf}

\begin{document}

\title{Resolving spin currents and spin densities generated by charge-spin interconversion in systems with reduced crystal symmetry}

\author{Lorenzo Camosi}
\affiliation{Catalan Institute of Nanoscience and Nanotechnology (ICN2), CSIC and The Barcelona Institute of Science and Technology (BIST), Campus UAB, Bellaterra, 08193 Barcelona, Spain}
\author{Josef Sv\v{e}tlik}
\affiliation{Catalan Institute of Nanoscience and Nanotechnology (ICN2), CSIC and The Barcelona Institute of Science and Technology (BIST), Campus UAB, Bellaterra, 08193 Barcelona, Spain}
\affiliation{Universitat Aut\`{o}noma de Barcelona, Bellaterra, 08193 Barcelona, Spain}
\author{Marius V. Costache}
\affiliation{Catalan Institute of Nanoscience and Nanotechnology (ICN2), CSIC and The Barcelona Institute of Science and Technology (BIST), Campus UAB, Bellaterra, 08193 Barcelona, Spain}
\author{Williams Savero Torres}
\affiliation{Catalan Institute of Nanoscience and Nanotechnology (ICN2), CSIC and The Barcelona Institute of Science and Technology (BIST), Campus UAB, Bellaterra, 08193 Barcelona, Spain}
\author{Iv\'{a}n Fern\'{a}ndez Aguirre }
\affiliation{Catalan Institute of Nanoscience and Nanotechnology (ICN2), CSIC and The Barcelona Institute of Science and Technology (BIST), Campus UAB, Bellaterra, 08193 Barcelona, Spain}
\affiliation{Universitat Aut\`{o}noma de Barcelona, Bellaterra, 08193 Barcelona, Spain}
\author{Vera Marinova}
\affiliation{Institute of Optical Materials and Technologies, Bulgarian Academy of Science, Sofia 1113, Bulgaria}
\author{Dimitre Dimitrov}
\affiliation{Institute of Optical Materials and Technologies, Bulgarian Academy of Science, Sofia 1113, Bulgaria}
\affiliation{Institute of Solid State Physics, Bulgarian Academy of Sciences, 1784 Sofia, Bulgaria}
\author{Marin Gospodinov}
\affiliation{Institute of Solid State Physics, Bulgarian Academy of Sciences, 1784 Sofia, Bulgaria}
\author{Juan F. Sierra}
\affiliation{Catalan Institute of Nanoscience and Nanotechnology (ICN2), CSIC and The Barcelona Institute of Science and Technology (BIST), Campus UAB, Bellaterra, 08193 Barcelona, Spain}
\author{Sergio O. Valenzuela}
\affiliation{Catalan Institute of Nanoscience and Nanotechnology (ICN2), CSIC and The Barcelona Institute of Science and Technology (BIST), Campus UAB, Bellaterra, 08193 Barcelona, Spain}
\affiliation{Instituci\'{o} Catalana de Recerca i Estudis Avan\c{c}ats (ICREA), 08010 Barcelona, Spain}

\begin{abstract}
The ability to control the generation of spins in arbitrary directions is a long-sought goal in spintronics. Charge to spin interconversion (CSI) phenomena depend strongly on symmetry. Systems with reduced crystal symmetry allow anisotropic CSI with unconventional components, where charge and spin currents and the spin polarization are not mutually perpendicular to each other. Here, we demonstrate experimentally that the CSI in graphene-WTe$_2$ induces spins with components in all three spatial directions. By performing multi-terminal nonlocal spin precession experiments, with specific magnetic fields orientations, we discuss how to disentangle the CSI from the spin Hall and inverse spin galvanic effects.
\\

\end{abstract}

\maketitle

\section{INTRODUCTION}

In condensed matter, spin orbit coupling (SOC) and (broken) crystal and temporal symmetries play a fundamental role, strongly modifying the electronic states and connecting spin and orbital angular momentum degrees of freedom \citep{elliott1954spin}. Their action leads to novel physical states, such as topological phases \citep{hasan2010colloquium,bogdanov2001chiral,kane2005z,chang2016quantum}, and technologically relevant electron-spin transport phenomena, such as charge-spin interconversion (CSI) \citep{manchon2015new,sinova2015spin,soumyanarayanan2016emergent,regina2021}.
The spin Hall effect (SHE) \citep{sinova2015spin,d1971possibility,dyakonov1971current,hirsch1999spin} and inverse spin galvanic effect (ISGE) \citep{ganichev2002spin} (and the corresponding reciprocal effects according to the Onsager relationships \citep{onsager1931reciprocal,jacquod2012onsager}) are fundamental CSI phenomena that have been broadly investigated as spin generators and detectors \citep{manchon2015new,sinova2015spin,soumyanarayanan2016emergent}. In the conventional SHE, an electrical current induces a transverse spin current. The basic mechanism can have extrinsic or intrinsic origin \citep{sinova2015spin}; the former involves Mott scattering with impurities, while the latter is closely connected to the Berry curvature \citep{berry1984quantal,xiao2010berry,chang2016quantum}. In the ISGE, also known as Rashba-Edelstein effect \citep{edelstein1990spin}, an electrical current induces a non-equilibrium spin density. The ISGE results from a redistribution of charge carriers on the Fermi surface in systems having a momentum-asymmetric spin texture, which derives from a broken inversion symmetry, either structural (\textit{e.g.} surface or interface) or in the bulk (\textit{i.e} crystal lattice) \citep{luo2009full,zhang2014hidden}.

Experimental observations of the SHE and ISGE were originally obtained in semiconductors \citep{ganichev2002spin,kato2004observation,wunderlich2005experimental} and metals \citep{valenzuela2006direct,saitoh2006conversion,kimura2007room}. Even though it was known that, given their SOC-related origin, the SHE and ISGE are often concomitant, those early works usually focused on either the SHE or the ISGE (and their reciprocals). However, understanding the relation between the SHE and ISGE has become essential in light of their potential technological relevance, in particular for electrically reorienting magnets for memory applications \citep{sinova2015spin,dieny2020natelect}. Research on SHE and ISGE has been further stimulated by recent results in van der Waals heterostructures \citep{JFS2021NatNano} and, more specifically, in graphene in proximity with high-SOC materials, in which the CSI efficiency is found to be relatively large \citep{ghiasi2019charge,safeer2019room,benitez2020tunable}.

In some cases, it is possible to discriminate between the SHE and ISGE as, for instance, when graphene is modified by the proximity of a semiconducting transition metal dichalcogenide (TMDC) in high-symmetry heterostructures. There, the SHE and ISGE are driven by valley-Zeeman and conventional Rashba SOC, respectively, leading to spin populations that are orthogonal to each other \cite{jose2017,offidani2017} and, therefore, that can be easily disentangled \citep{ghiasi2019charge,benitez2020tunable,cavill2020proposal}. This is not the case when the TMDC is conducting, as separating contributions deriving from currents in the TMDC bulk and in the interface or in proximitized graphene is not straightforward \citep{safeer2019room,JFS2021NatNano}. In addition, if the TMDC is a low-symmetry material, such as MoTe$_2$ or WTe$_2$, or twisting between graphene and the TMDC results in an heterostructure with reduced-symmetry, CSI with unconventional spin orientations can be expected \citep{safeer2019large,li2019twist,zhao2020unconventional,naimer2021}.

Spin-torque experiments have proposed the presence of unconventional torques compatible with the WTe$_2$ symmetries \cite{macneil2017nphys}. However, recent studies in graphene-MoTe$_2$ and graphene-WTe$_2$ nonlocal spin devices, not only have not observed the new CSI contribution but found instead a component of unknown origin \citep{safeer2019large,zhao2020unconventional}, which was ascribed to broken symmetries due to uncontrolled strain during device fabrication. These measurements were carried out without characterization of the strain or direct knowledge of the crystal orientation, although it has been argued that the crystals typically cleave in a favored direction \citep{safeer2019large}. Furthermore, a recent study suggests that a widespread method to identify the ISGE (SGE) by rotating the magnetization of the ferromagnetic (FM) detector (injector) is unreliable \citep{safeer20212dmater}. These observations demonstrate that it is necessary to carry out additional experiments and to establish nonlocal measurement schemes to identify and quantify competing CSI effects in combination with crystal-orientation and strain characterization in the same device.

In this work, we present a measurement protocol based on spin precession for resolving the contributions of SHE and ISGE \citep{benitez2020tunable,safeer20212dmater}.
We then implement a graphene-WTe$_2$ device and demonstrate the generation of spins oriented within and perpendicular to the graphene plane. Although an out-of-plane spin polarization by CSI is expected due to the crystal symmetry in thin WTe$_2$ layers, it has never been observed in hybrid graphene-WTe$_2$ heterostructures. Here, we detect it for the first time, demonstrating that this system enables spin generation with all three spatial orientations. The WTe$_2$ crystal orientation is determined by means of polarized Raman spectroscopy, which also characterizes the crystal strain. Based on these observations, the possible origins of the observed CSI components are discussed.

\section{RESULTS AND DISCUSSION}

\textbf{CSI and crystal symmetries}. The character of the CSI is dictated by crystal symmetries. The SHE is described by linear response theory \citep{seemann2015symmetry} as $j_i^k = \sigma^{\textit{k}}_{\textit{ij}}E_j$, where $E_j$ is the external electric field in the $\hat{j}$ direction that generates a charge current $\mathbf{j}_c$, $j_i^k$ the spin current in the $\hat{i}$ direction with spin polarization $\mathbf{s}$ in the $\hat{k}$ direction and $\sigma^{\textit{k}}_{\textit{ij}}$ the spin conductivity tensor. In a high-symmetry crystal only off-diagonal $\sigma^{\textit{k}}_{\textit{ij}}$ terms with $ i\neq j\neq k$ are non-zero, resulting in ($\mathbf{j}_c \perp \mathbf{j}_s \perp \mathbf{s}$). Reduced symmetries allow additional $\sigma^{\textit{k}}_{\textit{ij}}$ elements that can break the mutual perpendicular relationship between $\mathbf{j}_c$, $ \mathbf{j}_s $ and $ \mathbf{s}$ \citep{seemann2015symmetry,roy2022}. For example, in a crystal with a single mirror plane, as depicted in Fig. 1a, a $\mathbf{j}_c$ perpendicular to the plane can lead to $ \mathbf{s}$ parallel to the transverse $ \mathbf{j}_s $ (Fig. 1b) but mirror symmetry still precludes a transverse $ \mathbf{j}_s $ with an $ \mathbf{s}$ component along $\mathbf{j}_c$. The latter restriction disappears if the mirror symmetry is broken. The ISGE depends directly on the electronic band structure polarization but it is governed by the same symmetry considerations \citep{luo2009full,zhang2014hidden}. In a high symmetry crystal, non-zero spin density could only arise at a boundary or interface, leading to the conventional Rashba effect \citep{bychkov1984oscillatory,bihlmayer2015focus}. In Fig. 1a, the mirror symmetry imposes that the spin polarization must be contained in the mirror plane (Fig. 1c).

\begin{figure}[h!]
\centering
\includegraphics[width=14cm]{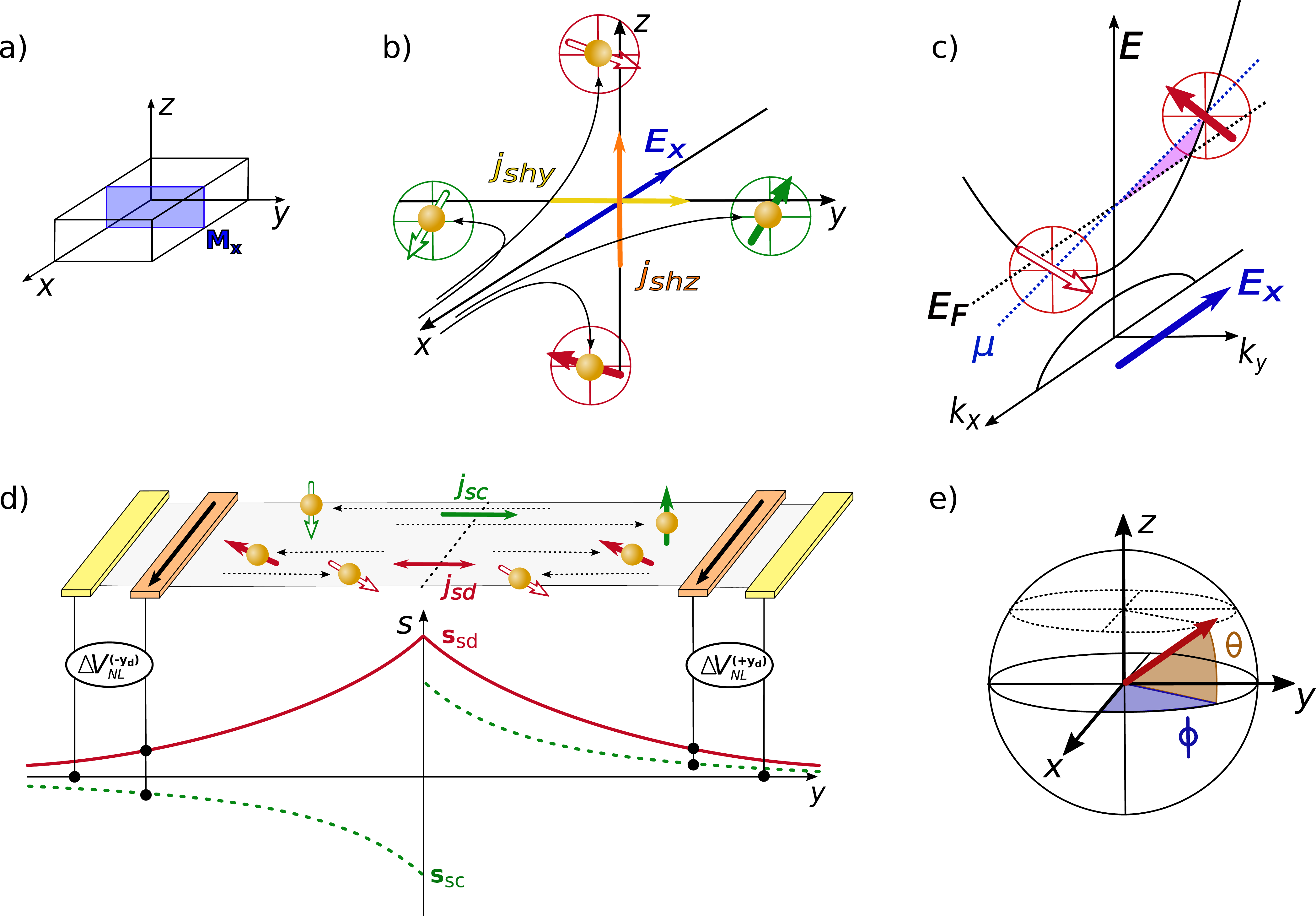}
\caption{Charge spin interconversion (CSI) in low symmetry structures. a) Representation of a crystal structure with a mirror plane perpendicular to $\hat{x}$ and allowed spin polarizations in the spin Hall (b) and the inverse spin galvanic (c) effects (SHE, ISGE). The electric field $E_x$ (or charge current) is applied in the $\hat{x}$ direction.
The spin polarizations ($\mathbf{s}_i$) are represented with arrows. The circles show the planes where spin polarization is allowed by symmetry. The spin Hall current in b) has components in $\hat{z}$ ($J_{shz}$) and $\hat{y}$ ($J_{shy}$). In c) a sketch of a parabolic band structure shows the out of the equilibrium electron population distribution under $E_x$. d) Sketch of the device geometry and generated spin currents (top) and spin electrochemical potential $s$ (bottom). The high-SOC material region at the center of the diagram, where the CSI occurs owing to a current in $\hat{x}$, is not shown for clarity. The spins generated in the CSI region diffuse in the spin channel towards $\pm y$ and are detected by the FM electrodes (orange). The spin currents $J_{sc}$ and $J_{sd}$ originate, respectively, from the spin Hall current $J_{shy}$ (panel b) and the uniform spin accumulation in the CSI region associated to $J_{shz}$ (panel b) and the spin density due to ISGE (panel c). Because of the $J_{sc}$ contribution, the orientation of the spins flowing towards the right ($+$) and left ($-$) of the CSI region can be different. e) Representation of the CSI spin polarization orientation with angles $\phi$ and $\theta$.}
\end{figure}

\textbf{Resolving SHE and ISGE in nonlocal devices}. Nonlocal spin-dependent measurements have been widely used to investigate CSI phenomena \citep{valenzuela2006direct} and extract the polarization of the generated spins. There, a charge current $I$ is applied along the CSI region and the nonlocal voltage $\Delta V_{NL}$ is measured between a FM detector and a reference outer metallic contact as a function of the orientation and magnitude of an applied magnetic field $\mathbf{B}$ \citep{valenzuela2006direct,WST2017,yan2017}. When the spins originating from the SHE and the ISGE are perpendicular to each other, they can be fully resolved by investigating spin precession with in-plane and perpendicular magnetic fields \citep{benitez2020tunable} or by means of a symmetry analysis under oblique magnetic fields \citep{ghiasi2019charge,cavill2020proposal}. These approaches are not sufficient in graphene-WTe$_2$ or graphene-MoTe$_2$ heterostructures, where the TMDC is conducting and has low-symmetry \citep{zhao2020unconventional,zhao2020observation,safeer2019large}. However, in such systems the spin density induced by the ISGE has no preferential direction, whereas the SHE induces a directed spin current. Therefore, it is possible to discern between the SHE and ISGE generated spins by simultaneously measuring the spin accumulation along the spin-current direction at opposite sides of the CSI region, as illustrated in Fig. 1d.

When the current $I$ is applied along $\hat{x}$ in the CSI region (not shown in the Fig. 1d), the SHE creates spin currents along $\hat{y}$ ($J_{shy}$) and $\hat{z}$ ($J_{shz}$) (Fig. 1b), which coexist with the spin density induced by the ISGE (Fig. 1c). A direct comparison between $\Delta V_{NL}$ at $\pm y$, $\Delta V_{NL}^{(\pm y)}$, differentiates the spins associated to the spin current $J_{shy}$ from those associated to the ISGE and $J_{shz}$ (Fig. 1d). Indeed, the spin current component $J_{sc}$ diffusing away from the CSI region, which originates from $J_{shy}$, generates opposite spin accumulation at $\pm y$, while the spin current $J_{sd}$, associated to the ISGE and $J_{shz}$, generates equal spin accumulation at $\pm y$. Therefore, $(\Delta V_{NL}^{(+y)} + \Delta V_{NL}^{(-y)})/2 \propto J_{sd}$ whereas  $(\Delta V_{NL}^{(+y)} - \Delta V_{NL}^{(-y)})/2 \propto J_{sc}$. Such measurements do not distinguish between the ISGE and $J_{shz}$, since both induce a spin density whose orientation does not vary in the CSI region \citep{safeer2019room}. Nevertheless, these components can in principle be disentangled by analysing $(\Delta V_{NL}^{(+y)} + \Delta V_{NL}^{(-y)})/2 $ as a function of the SOC-material thickness \citep{sinova2015spin}. In particular, when the thickness is much smaller than the spin relaxation length along $\hat{z}$, $J_{shz}$ vanishes \cite{zhang2001prl,stamm2017prl}, and only ISGE contributes to the nonlocal signal.

With the previous considerations, we focus on fully characterizing the spins generated by $J_{sc}$ and $J_{sd}$ by means of spin precession using spin detectors located at $\pm y_d$. The steady-state spin diffusion and precession in the spin channel are governed by the Bloch diffusion equation \cite{torrey,js1988,zutic2004}:

\begin{equation}
\mathbf{D}\frac{\partial^2 \mathbf{s}}{\partial y^2} + \gamma_c \mathbf{s} \times \mathbf{B} - \mathbf{s} \cdot \overline{\overline{\Gamma}}^{-1} = 0
\end{equation}

\noindent where $\mathbf{s}=(s_x,s_y,s_z)$ and $s_i$ is the spin electrochemical potential for spins along $i$, $\gamma_c$ is electron gyromagnetic ratio, $\mathbf{D}=(D_x,D_y,D_z)$ is the diffusion constant, and $\overline{\overline{\Gamma}}$ characterizes the spin lifetime $\tau_i$. In the most general case, the injected spins have an arbitrary spin orientation. The CSI spin injection efficiency into the spin channel can be quantified using effective spin-polarization factors $\mathbf{P}= (P_x,P_y,P_z)= (j_x / I ,j_y / I ,j_z / I ) w$, with $j_i$ the corresponding spin currents with contributions from $J_{sc}$ and $J_{sd}$, and $w$ the width of the channel. Assuming isotropic spin transport in the channel and that the FM detector is characterized by a magnetization along the $+\hat{x}$ direction and by a polarization efficiency $P_{FM}$, the nonlocal resistance $R_{NL} \equiv \Delta V_{NL}/I$ for $\mathbf{B}$ along $\hat{y}$ and $\hat{z}$ takes the respective general forms:

\begingroup\makeatletter\def\f@size{8}\check@mathfonts
\def\maketag@@@#1{\hbox{\m@th\large\normalfont#1}}%
\begin{align}
R_{NL}(B_y)= \frac{R_{\Box} P_{FM} }{4 w} Re\left[ \frac{ i e^{-\sqrt{\frac{1-i \tau \omega }{D \tau}} y_d} ( P_x + i P_z)}{\sqrt{\frac{1-i \tau \omega }{D \tau}}} + h.c.\right]  \\
R_{NL}(B_z)= \frac{R_{\Box} P_{FM}}{4 w} Re\left[ \frac{ i e^{-\sqrt{\frac{1-i \tau \omega }{D \tau}} y_d} ( P_x - i P_y)}{\sqrt{\frac{1-i \tau \omega }{D \tau}}} + h.c.\right]
\end{align}\endgroup

\noindent where $\omega= \gamma_c B $ and $R_{\Box}$ is the channel sheet resistance.

Fitting the spin precession response to Eqs. (2) and (3) determines $\mathbf{P}$. The spin-injection angles for spins moving towards  $\pm \hat{y}$, defined as $ \theta^\pm \equiv \arctan ( P_z^{\pm} / \sqrt{(P_x^\pm)^2+(P_y^\pm)^2} )$ and $\phi^\pm  \equiv \arctan(P_y^\pm  / P_x^\pm )$, fully characterize the orientation of the injected spins on both sides of the CSI region (Fig. 1e). Therefore, comparing $\theta^+$ with $\theta^-$ and $\phi^+$ with $\phi^-$ provides direct information on $J_{sc}$ and $J_{sd}$. In addition, focusing on $\theta ^{\pm}$ and $\phi^{\pm}$, rather than $\mathbf{P^{\pm}}$, eliminates systematic errors deriving from potential differences in the FM detector polarizations.

\textbf{CSI in graphene-WTe$_2$}. WTe$_2$ is a layered TMDC, which is stable in the orthorhombic Td(Pmn2$_1$) phase. It is characterized by a lack of bulk inversion symmetry (Fig. 2a-b), a glide mirror plane $M_b$, an out-of-plane two-fold screw rotational symmetry $C_{2c}$ and a mirror plane $M_a$ perpendicular to the $\hat{a}$ crystallographic direction (Fig. 2b). Multilayer Td-WTe$_2$ is a type-II Weyl semimetal \citep{soluyanov2015type,li2017evidence} while in  monolayer form it is a 2D topological insulator \citep{fei2017edge,tang2017quantum,wu2018observation}. Considering the mirror crystal symmetries, in the SHE, $\mathbf{j}_c \perp \mathbf{j}_s \perp \mathbf{s}$ while, in the ISGE, $\mathbf{j}_c$ along $\hat{a}$ can induce a spin density with spins oriented along $\hat{b}$ and viceversa, but no spin density parallel to $\hat{c}$.

The graphene-WTe$_2$ device fabrication follows the protocols established with other TMDCs \citep{benitez2020tunable} (see Supplementary Information \cite{SI}). The WTe$_2$ crystals were grown by chemical vapor transport, using bromine as a transport agent \citep{dimitrov2020chemical}. Their electronic and crystalline structure was investigated by x-ray photoelectron spectroscopy, angle-resolved photoemission spectroscopy, and Raman spectroscopy, all of them demonstrating high quality and confirming the Td phase \citep{SI}. The crystallographic orientation of the WTe$_2$ crystal in the actual device (Fig. 2c) is obtained by means of linearly polarized Raman spectroscopy. Figure 3d shows the Raman spectra when the laser polarization is rotated an angle $\alpha$ relative to the long-axis of the WTe$_2$ flake (Fig. 2c). The intensity of the Raman modes A$^5_1$ (164 cm$^{-1}$) and A$^2_1$ (212 cm$^{-1}$) are known to change as a function of crystal orientation. As the ratio A$^5_1$/A$^2_1$ is maximum at $\alpha=0$ (Fig. 2e) then, in our device, $\hat{a}\parallel \hat{x}$ \citep{song2016scirep}.

As the spin transport in graphene is isotropic \cite{Raes2016}, the parameters $\tau$ and $D$ (as well as $P_{FM}$) in Eq (2) and (3) are extracted by measuring spin precession with out-of-plane $\mathbf{B}$ in the two reference graphene devices at both sides of WTe$_2$. The CSI is then investigated by applying a charge current $I$ along the WTe$_2$ crystal ($\hat{a}\parallel \hat{x}$ direction). All measurements are carried out at room temperature.

\begin{figure}[h]
\centering
\includegraphics[width=16cm]{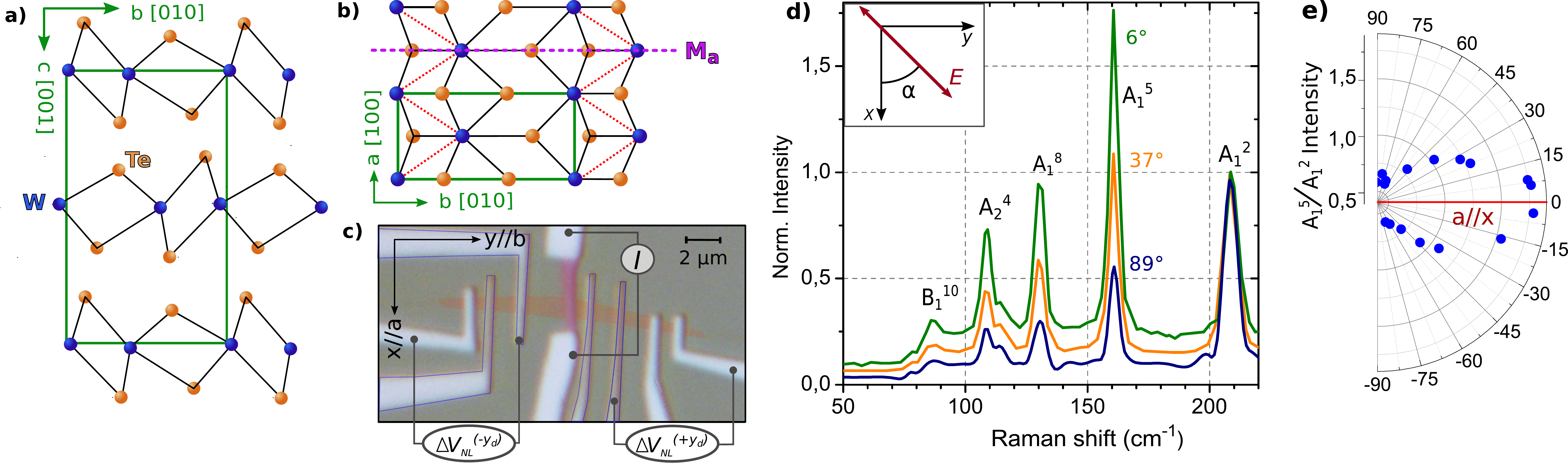}
\caption{ a,b) Crystal structure of WTe$_2$ in the Td phase. The W (Te) atoms and the atomic bonds are represented by blue (orange) circles and black lines, respectively. Green lines enclose the crystal unit cell. a) and b) represent the view along the crystal $a$ and $c$ axes, respectively. The mirror plane perpendicular to $a$, Ma, is shown in b). c) Optical image of the measured device. The graphene flake is highlighted in orange whereas the 8-nm thick WTe$_2$ has magenta contrast on the SiO$_x$(440nm)/Si substrate. The ferromagnetic contacts are bordered with a blue line. For the CSI measurements, a current $I = 7.5$ $\mu$A is applied along the WTe$_2$ flake while measuring the nonlocal voltages $\Delta V_{NL}^{(\pm y)}$. The distance of the FM detectors from the WTe$_2$ crystal is $y_d = 3$ $\mu$m. d) Polarized Raman spectra as a function of the WTe$_2$ flake orientation with respect to the laser polarization, given by the angle $\alpha$ (inset). The spectra are normalized with the intensity of the A$_1^2$ peak. e) Polar plot of the intensity ratio A$_1^5$/A$_1^2$ vs $\alpha$. The red line indicates the maximum of the A$_1^5$/A$_1^2$  intensity, which corresponds to the $\hat{a}$ crystallographic direction.}
\end{figure}

Figure 3 shows the nonlocal resistances $R_{NL}^{(+y_d)}$ and $R_{NL}^{(-y_d)}$ as a function of $B_y$ and $B_z$. The signals have been acquired with the FM magnetizations saturated along both the $\pm\hat{x}$ directions to remove contributions that are unrelated to spin and associated to the magnetization rotation of the FM detector \citep{SI}. Remarkably, $R_{NL}^{(+y_d)}$ and $R_{NL}^{(-y_d)}$ present nearly undistinguishable lineshapes for $B_z$ (Figs. 3a and 3b), while this is clearly not the case for $B_y$ (Figs. 3c and 3d).

For $B_z$, only the in-plane ($xy$) components of the injected spins contribute to the precession lineshape. Therefore, the fact that $R_{NL}^{(+y_d)} \sim R_{NL}^{(-y_d)}$ demonstrates that the spins difussing towards $-\hat{y}$ and $+\hat{y}$ have the same in-plane spin polarization, which is an indication of a uniform in-plane polarization in the CSI region (associated to $J_{sd}$ in Fig. 1d). Moreover, because $R_{NL}^{(\pm y_d)}$ are neither fully symmetric nor fully antisymmetric about $B_z = 0$, the spin polarization has nonzero components along both $\hat{x}$ and $\hat{y}$.

In contrast, for $B_y$ only the spin components in the $xz$ plane contribute to the precession lineshape. The marked difference between $R_{NL}^{(+y_d)}$ and $R_{NL}^{(-y_d)}$ demonstrates that the spins difussing towards $-\hat{y}$ and $+\hat{y}$ have different spin polarization orientation in the $xz$ plane. Combined with the results for $B_z$, this observation is an unambiguous indication of a spin polarized current in the CSI region with a polarization along $\hat{z}$ (associated to $J_{sc}$ in Fig. 1d). Furthermore, $R_{NL}^{(+y_d)}$ being rather symmetric about $B_y = 0$ (Fig. 3d) also demonstrates the presence of a uniform spin density along $\hat{z}$, originating from $J_{sd}$, which partially compensates the contribution from $J_{sc}$.

To quantify the relative magnitudes of each CSI component, we fit the measurements in Fig. 3 to Eqs. (2) or (3). The fittings are shown with blue lines, from which the spin-polarization angles $\theta ^{\pm}$ and $\phi^{\pm}$ are extracted: $(\phi^{-},\theta^{-})=(-40^{\circ} \pm 5 ^{\circ}, -34^{\circ} \pm 4 ^{\circ} )$ and $(\phi^{+},\theta^{+})=(-41^{\circ} \pm 3 ^{\circ},  -10^{\circ} \pm 6 ^{\circ} )$ (see Fig. 3e and 3f for a schematic representation). The spins originating from $J_{sd}$ and $J_{sc}$ relate to $(\phi_{s},\theta_{s})=[(\phi^+ + \phi^-)/2, (\theta^+ +\theta^-)/2] =(-41^{\circ} \pm 4^{\circ}, -22^{\circ} \pm 5^{\circ})$ and $(\phi_{as},\theta_{as})=[(\phi^+ -\phi^-)/2 , (\theta^+ -\theta^-)/2]=(0 \pm 4^{\circ},- 12^{\circ} \pm 5^{\circ})$, respectively. The presence of non-zero symmetric and antisymmetric angular components confirms the coexistence in the CSI region of spin-Hall currents along $\hat{y}$ with polarization in $\hat{z}$ and uniform spin densities with polarizations components on $\hat{x}$, $\hat{y}$ and $\hat{z}$.

\begin{figure}[h]
\centering
\includegraphics[width=10cm]{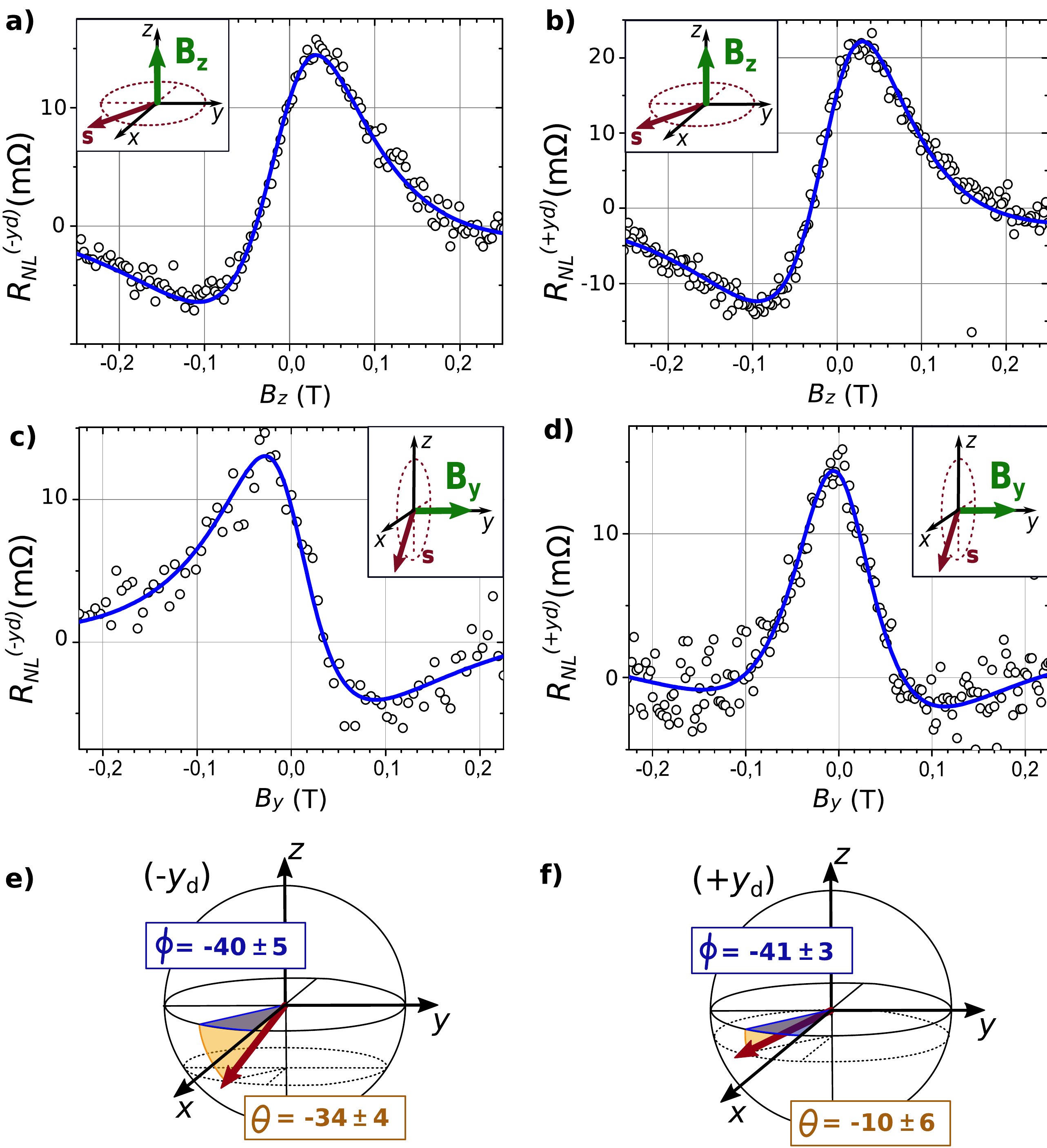}
\caption{Nonlocal resistance $R_{NL}^{(\pm y_d)}$ at $-y_d$ and $+y_d$ for magnetic field $B_z$ (a, b) and $B_y$ (c, d). Open circles show the experimental data and the blue lines the corresponding fit to Eqs. (2) or (3). The insets represent the magnetic field direction and the correspondent spin precession plane. The odd and even parts in the precession response for both $B_z$ and $B_y$ demonstrates that the spin polarization generated in the CSI region has components in all three axis. The coincidence in the measurements with $B_z$ shows that there is no net spin current flowing along $\hat{y}$ with polarization in the $xy$ plane. The stark difference in the measurements with $B_y$ highlights the presence of both a large spin density and spin current flowing along $\hat{y}$, with polarization in the $xz$ plane. Combining $B_z$ and $B_y$ results, the spin polarized current can only be polarized in $\hat{z}$. e) and f) CSI angles extracted from the fits of $R_{NL}^{(\pm y_d)}$.}
\end{figure}

\textbf{Origin of the observed CSI in graphene-WTe$_2$}. Although our experiments establish the presence of CSI with different symmetries, unequivocally identifying the underlying CSI mechanisms requires further analysis. The CSI can originate from either the ISGE or the SHE, in the bulk of WTe$_2$, at the graphene-WTe$_2$ interface or in graphene by proximity effects \citep{JFS2021NatNano}.

The spin density in the CSI region has components in all three directions (Fig. 3). According to the Td-WTe$_2$ crystal symmetries, for a current in $a$ ($\hat{x}$) only the spin-polarization component along $b$ ($\hat{y}$) is allowed: the ISGE can generate a spin density in $\hat{y}$, while the spin Hall current flowing in $\hat{z}$ should also be polarized along $\hat{y}$ \cite{xue2020staggered}. Therefore, the $\hat{x}$ and $\hat{z}$ components are not expected in the bulk of Td-WTe$_2$, which indicates that both originate from interfacial or proximity-induced effects in graphene considering the reduced symmetry of the heterostructure \cite{xue2020staggered}.

The symmetry of an heterostructure or thin crystal can be equal or lower than its bulk constituents, permitting additional non-zero spin polarization components. Indeed, in the graphene-WTe$_2$ interface, the glide symmetry is absent, leaving possibly only the single mirror symmetry plane $M_a$. As discussed in Fig. 1, the generation of a spin density in $c$ ($\hat{z}$), when $\mathbf{j}_c$ is along $a$ ($\hat{x}$), would then be allowed.

The spin density along $\hat{x}$ was previously observed in graphene-MoTe$_2$ and graphene-WTe$_2$ and was ascribed to the presence of an additional bulk mirror symmetry breaking induced by strain in the TMDC \citep{safeer2019large,zhao2020unconventional}. In our experiments, Raman spectroscopy demonstrates that WTe$_2$ is not under strain \citep{SI}, making this explanation implausible. Alternatively, the spin density in $\hat{x}$ could be generated by a current component along $\hat{y}$. Recent theoretical works reported anisotropic SHE in WTe$_2$ and MoTe$_2$ as a function of charge current direction and Fermi energy position \cite{zhou2019intrinsic,vila2021low}. However, the elongated geometry of our WTe$_2$ flake (Fig. 2d) implies that the current component along $\hat{y}$ is very small and a very large CSI efficiency would be required to make this scenario viable. These observations confirm that the spin density along $\hat{x}$ is likely generated in the proximitized graphene.

Recent first-principles calculations addressing the twist-angle dependence of proximity-induced SOC in graphene by TMDCs MoS$_2$, MoSe$_2$, WS$_2$ and WSe$_2$ have shown that the Rashba SOC could exhibit a radial component, thus deviating from the typical tangential orientation \cite{naimer2021}. As discussed in refs. \cite{li2019twist,naimer2021}, the radial component is allowed for twist angles between the graphene and TMDC lattices different from 0$^\circ$ and 30$^\circ$, where mirror symmetries are broken and the point group symmetry of graphene-TMDC is reduced to $C_3$. Similarly, an arbitrary twist angle between the graphene and WTe$_2$ crystal lattices can break the remaining symmetry upon reflection in $M_a$. At their interface, with radial Rashba SOC coupling, a spin polarization component parallel to $\mathbf{j}_c$ can then arise.

It remains to be understood why the radial component has never been observed in graphene-WS$_2$ and graphene-MoS$_2$ \cite{ghiasi2019charge,safeer2019room,benitez2020tunable}. The high annealing temperatures used in those works, to clean the heterostructures  and improve the interface quality, might have favored 0$^\circ$ or 30$^\circ$ twist angles, although this argument would not be valid if the initial (arbitrary) twist angle was larger than a few degrees. Therefore, further studies are required to address this question.

Finally, the observed spin current along $\hat{y}$, which is polarized in $\hat{z}$, has not been reported in prior graphene-WTe$_2$ studies \citep{zhao2020unconventional,zhao2020observation}. It is however allowed even in high-symmetry structures, and has been reported in graphene-MoS$_2$ and graphene-WS$_2$ \cite{ghiasi2019charge,safeer2019room,benitez2020tunable}. In these experiments, the generation of a spin density with polarization in $\hat{y}$ was also found. It was further confirmed that both the spin current polarized in $\hat{z}$ and the spin density in $\hat{y}$ can originate solely from the SHE and the ISGE in proximitized graphene, respectively \cite{ghiasi2019charge,benitez2020tunable}. But, in general, the former could be due to the SHE, in graphene or in the TMDC, while the latter to the ISGE in graphene or the SHE in the TMDC \cite{safeer2019room}.

The previous discussion strongly suggests that the unconventional CSI components in graphene-WTe$_2$ originate from interfacial or proximity effects. In contrast, the observed spin current with polarization in $\hat{z}$ and the spin densities in $\hat{y}$ are permitted both in the bulk of Td-WTe$_2$ and in proximitized graphene. Quantifying the spin absorption in WTe$_2$ could in principle help separate these remaining CSI contributions, however, the analysis is not straightforward or free of ambiguities. The estimation of the spin absorption requires detailed knowledge of heterostructure properties that cannot be readily obtained in nonlocal devices. The properties include the precise interface resistance between graphene and WTe$_2$ as well as the spin relaxation parameters in both the proximitized graphene and WTe$_2$. Any subtle change in these parameters, or in the implementation of the spin absorption model, can result in diverging conclusions. In addition, due to the 2D nature of graphene, the spin absorption is not uniform at the graphene-WTe$_2$ interface, even if the spin-current absorption occurs in $z$ direction. This can be easily understood by considering the inverse SHE, commonly used in CSI experiments. Because of spin relaxation and the fact that there is no alternative path for spins to cross the CSI region, as in a 3D system, the majority of spins will be absorbed on the side of graphene-WTe$_2$ that is closest to the FM injector. This leads to a spin accumulation gradient (and a spin current) in the TMDC along the spin channel.

\section{CONCLUSIONS}

We have demonstrated experimentally that the CSI in graphene-WTe$_2$, with a current along the WTe$_2$ $a$ axis, induces spin-polarized carriers with polarization components in all three spatial directions. By implementing systematic multi-terminal nonlocal spin precession experiments, we have shown that it is possible to disentangle the CSI from the spin Hall and inverse spin galvanic effects. A spin current flowing along the graphene channel leads to opposite spin polarization on the two sides of the CSI region, inducing a signal with opposite sign in remote FM detectors at each side. In contrast, a spin density in the CSI region produces a signal with equal sign in the same detectors.

We confirmed a spin accumulation with polarization along the applied current, which by symmetry is not allowed in the WTe$_2$ bulk. Our analysis indicates that this spin accumulation originates at the graphene-WTe$_2$ interface and involves the emergence of a radial component in the proximity-induced Rashba SOC, which arises from twisting \cite{li2019twist,naimer2021}. To validate this interpretation, additional experiments are needed. In particular, the predicted radial component was predicted to be extremely sensitive to the twist angle, unintended doping and carrier density \cite{naimer2021}. Therefore, it is necessary to systematically address the dependence of the spin accumulation parallel to the current as a function of crystalline orientation, sample annealing temperature and gate voltage. For the latter, thin, ideally monolayer, WTe$_2$ should be used to avoid both gate shielding (from semi-metallic bulk WTe$_2$) and artifacts from spin absorption. Nevertheless, to unambiguously demonstrate the radial Rashba component, it might be better to simply focus on semiconducting TMDCs.

We also show, for the first time, the presence of spin currents and spin densities with polarization perpendicular to the substrate plane in graphene-WTe$_2$. The observed spin polarization, as generated by the SHE spin current is always allowed, and can be induced in the bulk of WTe$_2$ or in proximitized graphene. However, symmetry considerations show that the spin density with perpendicular polarization must originate from the ISGE at the graphene-WTe$_2$ interface.

Overall, our work demonstrates that multi-terminal CSI measurements combined with symmetry analysis is a powerful approach to discriminate CSI signals. This is important for both applied and fundamental reasons. In particular, understanding the CSI originating from SHE and ISGE with all possible orientations is a promising route for magnetic memory applications, where the generation of unconventional spin-orbit torques is required \cite{macneil2017nphys}.

\section{ACKNOWLEDGEMENTS}
We acknowledge support of the European Union's Horizon 2020 FET-PROACTIVE project TOCHA under grant agreement 824140 and of the Spanish Research Agency (AEI), Ministry of Science and Innovation, under contracts no. PID2019-111773RB-I00/AEI/10.13039/501100011033, and SEV-2017-0706 Severo Ochoa. J. F. S acknowledges support from AEIunder contract RYC2019-028368-I/AEI/10.13039/50110001103, W. S. T from the European Union Horizon 2020 research and innovation program, grant number 881603 (Graphene Flagship), and I. F. A. of a fellowship from ”la Caixa” Foundation (ID 100010434) with code LCF/BQ/DI18/11660030 and of H2020 Marie Skłodowska-Curie grant agreement No. 713673.

\bibliographystyle{unsrt}
\bibliography{Bib}

\end{document}